\documentclass[aps,superscriptaddress,eqsecnum,nofootinbib,showpacs,preprintnumbers]{revtex4-2}
\usepackage[utf8]{inputenc}
\usepackage{amsmath}
\usepackage{amssymb}
\usepackage{amsfonts,amsthm,bm}
\usepackage{color,xcolor}
\usepackage{comment}
\usepackage{soul}

\usepackage{graphicx,epsfig}
\usepackage{float}
\usepackage{subcaption}
\usepackage{tikz}
\newcommand{\be}{\begin{eqnarray}}
	\newcommand{\ee}{\end{eqnarray}}
\newcommand{\bea}{\begin{eqnarray}}
	
	\newcommand{\eea}{\end{eqnarray}}

\newcommand*\diff{\mathop{}\!\mathrm{d}}

\usepackage{ulem}

\def\del{\partial}

\definecolor{azure(colorwheel)}{rgb}{0.0, 0.5, 1.0}
\definecolor{DarkViolet}{RGB}{148,0,211}
\definecolor{myDarkBlue}{rgb}{0,0.1,0.7}
\definecolor{DarkBlue}{RGB}{0,0,153}
\definecolor{amber}{rgb}{1.0, 0.49, 0.0}
\definecolor{amaranth}{rgb}{0.9, 0.17, 0.31}
\definecolor{nicered}{rgb}{0.7,0.1,0.1}
\definecolor{brown}{rgb}{0.5,0.1,0.1}
\definecolor{nicegreen}{rgb}{0.0,0.3,0.0}
\definecolor{tealgreen}{rgb}{0.0, 0.51, 0.5}

\definecolor{tclr}{RGB}{148,0,211}

\usepackage{hyperref}
\hypersetup{colorlinks,bookmarksopen,
	bookmarksnumbered,
	citecolor={nicered},
	linkcolor={myDarkBlue},
	urlcolor={tealgreen},
	pdfstartview=FitH}

\newcommand{\beq}{\begin{equation}}
\newcommand{\eeq}{\end{equation}}
\newcommand{\bseq}{\begin{subequations}}
	\newcommand{\eseq}{\end{subequations}}

\usepackage{times}

\graphicspath{{Figs/}}

\usepackage{orcidlink}
\def\idbako{\orcidlink{0000-0002-3012-6144}}
\def\idchat{\orcidlink{0000-0003-4479-2970}}
\def\idkara{\orcidlink{0000-0002-5479-6513}}

\begin{document}
\setcounter{page}{1}
\title[]{Thermodynamics of black holes featuring primary scalar hair}

\author{Athanasios Bakopoulos\idbako}
\email{atbakopoulos@gmail.com}
\affiliation{Physics Department, School of Applied mathematical and Physical Sciences,
	National Technical University of Athens, 15780 Zografou Campus,
	Athens, Greece.}

\author{Nikos Chatzifotis\idchat }
\email{chatzifotisn@gmail.com}
\affiliation{Physics Department, School of Applied mathematical and Physical Sciences,
	National Technical University of Athens, 15780 Zografou Campus,
	Athens, Greece.}

\author{Thanasis Karakasis\idkara}
\email{thanasiskarakasis@mail.ntua.gr}
\affiliation{Physics Department, School of Applied mathematical and Physical Sciences,
	National Technical University of Athens, 15780 Zografou Campus,
	Athens, Greece.}

\begin{abstract}
In this work, we embark on the thermodynamic investigation concerning a family of primary charged black holes within the context of shift and parity symmetric Beyond Horndeski gravity. Employing the Euclidean approach, we derive the functional expression for the free energy and derive the first thermodynamic law, offering a methodology to address the challenge of extracting the thermal quantities in shift-symmetric scalar tensor theories characterized by linear time dependence in the scalar field. Following the formal analysis, we provide some illustrative examples focusing on the thermal evaporation of these objects.
\end{abstract}
\maketitle

\flushbottom


\section{Introduction}\label{intro}
Horndeski gravity \cite{Horndeski:1974wa} stands as the most comprehensive scalar-tensor theory capable of circumventing Ostrogradsky instabilities. It establishes a mathematical framework, suitable for testing potential modifications of General Relativity through both minimal and non-minimal extensions of the action, which involve a single scalar degree of freedom. Although Horndeski theory was long believed to be the most general framework of scalar-tensor gravities, recent advancements \cite{Gleyzes:2014dya, Crisostomi:2016tcp, Kobayashi:2019hrl, Langlois:2015cwa,Langlois:2017mdk,Langlois:2018dxi}, have broadened the scope into the Beyond Horndeski and Degenerate Higher Order Scalar Tensor (DHOST) classifications. Notably, these theories adhere to degeneracy conditions to preclude the emergence of ghosts, while presenting an enriched class of effective field theories.

All these theories are well-known for their intricate nature. The inclusion of scalar field contributions within the action forms a theoretical framework that presents a highly challenging terrain for extracting non-trivial local solutions. A significant advancement in the realm of local solutions within shift and parity symmetric (Beyond) Horndeski gravities was recently achieved with the derivation of the first black holes featuring primary scalar charge\footnote{Primary charge is a new independent conserved quantity, while secondary charge is a quantity that depends on the initial charges of the spacetime.} \cite{Bakopoulos:2023fmv,Baake:2023zsq,Bakopoulos:2023sdm}. Primary scalar hair or charge can be understood as the physical attribute that distinguishes the black hole, alongside its mass, angular momentum, and electric charge. In this context, the primary charge is directly associated with the shift symmetry of the theory and is generated as the conserved charge of the corresponding Noether current. A key ingredient for the derivation of primary hair black holes is the existence of a linear time dependence in the ansatz of the scalar field, $\Phi=q t +\psi(r)$, which is permitted by the shift symmetry of the action, while the staticity of the metric remains unaffected \cite{Babichev:2013cya,Kobayashi:2014eva,Babichev:2015rva,Babichev:2017guv,Anson:2020trg,Charmousis:2021npl,BenAchour:2020fgy,Babichev:2024txe,Babichev:2024hjf,Takahashi:2019oxz,deRham:2019gha,Danarianto:2024vih}. Although the first solutions with linear time-dependent scalar appeared ten years ago, the investigation of the thermal properties of these configurations have remained an open issue \cite{Bravo-Gaete:2014haa}.

The exploration of black hole thermodynamics has emerged as a significant and captivating field, shedding light on the fundamental characteristics of black holes, while offering insights into the realm of quantum gravity \cite{Wald:1999vt,Carlip:2014pma,Jacobson:1995ab,Altamirano:2014tva}. This significance is particularly pronounced in the context of black holes within anti-de Sitter (AdS) space, where physical phenomena are correlated through the AdS/CFT duality \cite{Witten:1998zw}.
In this work, we derive the thermodynamics for a large class of asymptotically flat primary hair black hole solutions, tackling the non-trivial issue of the scalar field featuring linear time dependence.  An essential consideration pertains to assessing whether these configurations exhibit thermal stability, given the capacity of primary charge to influence the interior structure of black holes. This scrutiny is vital to ascertain their viability as potential remnants of the primordial universe. Additionally, since our solutions, in general, may have a cosmological constant, our analysis could be easily generalized to include (A)dS asymptotics and even derive their extended black hole thermodynamics \cite{Ahmed:2023snm,Xiao:2023lap}.

\section{Theoretical Framework}

We start our analysis with the most general  Beyond Horndeski action that respects both shift and parity symmetry under geometrized units ($c=G=1$),
\begin{equation}
    \label{eq:act}
    S= \int d^4x\frac{\sqrt{|g|}}{16\pi} \mathcal{L} = \int \frac{d^4x\sqrt{|g|}}{16\pi}\left[G_4(X)R+G_{4X}[(\square\Phi)^2-\Phi_{;\mu\nu}\Phi^{;\mu\nu}]+G_2(X)+F_4(X)\epsilon^{\mu\nu\rho\sigma}\epsilon^{\alpha\beta\gamma}_{\,\,\,\,\,\,\,\,\,\sigma}\Phi_{;\mu}\Phi_{;\alpha}\Phi_{;\nu\beta}\Phi_{;\rho\gamma}\right]\,,
\end{equation}
where for simplicity we denote the derivatives of the scalar field as $\Phi_{;\mu}\equiv\partial_\mu\Phi$, while  $\displaystyle X\equiv-1/2\partial^\mu\Phi\partial_\mu\Phi$ is the kinetic term of the scalar field. Regarding the scalar field, we adopt the following ansatz:
\begin{equation}
    \label{eq:phi}
    \Phi(t,r)=\chi(t)+\Psi(r)\,,\qquad \chi(t)=q t. 
\end{equation}
The linear time dependence in the expression of the scalar field $\Phi$ is allowed due to the shift symmetry of the considered Lagrangian density.
The internal shift symmetry of the theory \eqref{eq:act} results in the existence of a conserved Noether current, which is given by
\begin{equation}
    \label{eq:Noether}
   J^{\mu}=\frac{1}{\sqrt{|g|}}\frac{\delta S}{\delta (\del_\mu \Phi)}\,.
\end{equation}
 In a previous work, \cite{Bakopoulos:2023sdm}, we focused on static and spherically symmetric homogeneous black hole solutions with a primary charge emanating from the shift-symmetry of the scalar field. In particular, we use the following metric ansatz
\begin{equation}
    ds^2=-h(r)dt^2+\frac{dr^2}{h(r)}+r^2d\Omega^2.
\end{equation}
Following \cite{Bakopoulos:2023sdm}, it can be shown that the homogeneity of the solution is supported by the choice of $2XG_{4X}-G_4(X)+4X^2F_4(X)=-1$,
while the functionals $G_4(X)$ and $G_2(X)$ assume a linearly dependent form, i.e. $G_2=2b/\lambda^2 S(X)$ and $G_4(X)=1+b S(X)$. Regarding the action functional $S(X)$, we use the generic form
\begin{equation}
    \label{eq:S-exp}
    S(X)=\sum_{n=1}^\infty c_{\frac{n}{s}} X^{\frac{n}{s}}\,,\hspace{1em} s\in \mathbb{Z}^+\,,
\end{equation}
that has a smooth limit as $X\rightarrow 0$. The constant $s$ determines the step of the summation.  This will allow us to produce solutions in a semi-agnostic theory framework where all the information about the theory is encoded into the $c_i$ coefficients.

Although the theory is given in terms of an infinite power series, it leads to an integrable system that assumes the following solution
\begin{align}
    h=1-\frac{2M}{r}+\frac{\lambda b\sqrt{\pi}}{2r} \sum_{n=1}^\infty c_{\frac{n}{s}} \left(1-\frac{2n}{s} \right) \left(\frac{q^2}{2} \right)^{n/s} \frac{\Gamma\left(\frac{n}{s}-\frac{3}{2}\right)}{\Gamma\left(\frac{n}{s}\right)} 
    -\frac{2b}{3} \frac{r^2}{\lambda^2} \sum_{n=1}^{\infty} c_{\frac{n}{s}} \left(1-\frac{2n}{s} \right) \left(\frac{q^2}{2} \right)^{n/s} \,_2F_1\left(\frac{3}{2},\frac{n}{s};\frac{5}{2};-\frac{r^2}{\lambda^2} \right),
    \label{eq:h-hom-prop-F}
\end{align}
and
\begin{equation}
    \label{eq:X-hom-prop}
    X=\frac{q^2}{2}\frac{1}{1+(r/\lambda)^2}\,.
\end{equation}
The conservation of the shift-symmetry Noether current, $\nabla_\mu J^\mu=0$, 
results in the existence of a novel primary scalar charge which is related to the parameter $q$ and is given by
\begin{equation}
    \label{eq:sc-char-def}
    Q_s= \int  r^2 J^t\, \diff r=\frac{8\pi^{3/2}\beta}{N}\frac{\lambda}{q}\sum_{n=1}^\infty c_{\frac{n}{s}}\frac{n}{s}\left(\frac{q^2}{2}\right)^{n/s}\frac{\Gamma\left(\frac{n}{s}-\frac{3}{2}\right)}{\Gamma\left(\frac{n}{s}\right)}\,, \hspace{1em} \forall\,\frac{n}{s}>\frac{3}{2}\,,
\end{equation}
where $N$ is a normalization constant. 
By examining the properties of the above solution we note that the choice $s=2$, will always yield closed-form solutions for $n=2$, $n=4$, and any odd positive value of the integer $n$. 
Since the scalar charge is well-defined for $n/s>3/2$
for the rest of the article, we will be strictly focusing on the case of $s=2$ with $n\geq 4$.
We also deduce that these types of configurations can become regular. Indeed, by performing the expansion at $r=0$, we find that $ h(r)\sim1+a_0/r+a_1 r^2+\mathcal{O}(r^4)$, where $a_0$, and $a_1$ are the coefficients of the expansion. The case $a_0=0$, which makes the primary charge to secondary, 
 can regularize the black hole. See, for example, equation (\ref{regm}). This is only possible when $b\sum_{n=4}^{\infty}c_{n/2}<0$, which incidentally is also the condition for the preservation of the Weak Energy Condition, as it was verified in \cite{Bakopoulos:2023sdm}.

\section{Euclidean Thermodynamics}
We proceed with the discussion on the thermodynamics of black holes endowed with primary scalar hair. The black hole is enclosed within a cavity of a large radius, and we adopt the Grand Canonical Ensemble approach. Accordingly, the black hole is permitted to exchange energy with its environment, while maintaining a constant temperature $T$ and scalar voltage (or chemical potential for the scalar field) $\mathcal{W}$. Employing the ADM decomposition,
     \begin{equation}
        \label{lorentzmetric}
         ds^2=-N(r)^2 h(r)dt^2+\frac{dr^2}{h(r)}+r^2d\Omega^2
     \end{equation}
     and integrating out the angular component, we reach that our reduced action is expressed as
    
     \begin{equation}
        \label{reducedaction}
       S= \int dt \int d^3 x \frac{\sqrt{|g|}}{16\pi}\mathcal{L}=\int dt\int dr  \left\{\frac{N}{2}  \left(\frac{r^2+\lambda^2}{\lambda^2}+\frac{(\partial_t\chi)^2}{2 N^2 X}\right)b S- \frac{N}{2}  \left[\frac{(\partial_t\chi)^2 r X'}{N^2} (2 F_4)\right]-\frac{N}{2} \left(r h'(r)+h(r)-1\right)\right\},
     \end{equation}
where we made use of the constraint $\partial^2_t\chi=0$. The constraint is imposed by the shift symmetry, thereby maintaining the static nature of the metric. he imposition of the constraint in the above action is important for the correct calculation of the conjugate momentum. As the above action stands, it is easily verified that the effective degrees of freedom of our theory are $h,N, \chi, X$. Naturally, $X$ is dependent on $(\partial_t\chi)$ via
\begin{equation}
    \label{kinetic}
    X=-\frac{1}{2}\partial_\mu\Phi\partial^\mu\Phi=\frac{(\partial_t\chi)^2}{2 h N^2}-\frac{h (\Psi')^2}{2},
\end{equation}
however, the degree of freedom in $\Psi$ can be encoded in the kinetic term $X$ without any loss of information\footnote{Note that the use of $X$ instead of $\Psi$ is important for the correct calculation of the conjugate momentum.}, at least in the effective form of the Lagrangian that we have reached at this point. Therefore, we will be working in the subsequent calculations as though $\chi$ and $X$ are two \textit{independent} degrees of freedom of the reduced Lagrangian. Note that the function $\chi(t)$ is fixed via the constraint $\partial^2_t\chi=0$. However, its promotion to an effective degree of freedom is important for the correct definition of the conjugate momentum. This implies that the effective Lagrangian contains \textit{two effective scalar degrees of freedom}, in consistency with the scalar field decomposition of \eqref{eq:phi}. We have chosen such an approach because it simplifies the calculations tremendously, while the brute force approach is unmanageable due to the (in principle) unknown function $S(X)$ of our theory. The effective independence of $X$ and $\chi$ will be verified via the equations of motion where we will make use of the Euclidean approach to recover our solution with the Euclidean action, modulo the boundary terms, vanishing on-shell as expected.

 The Euclidean path integral in the saddle point approximation around the Euclidean solution is identified with the partition function of a thermodynamical ensemble \cite{Gibbons:1976ue}. Then, having obtained the on-shell value of the Euclidean action, denoted as $\mathcal{I}_E$, we will compare this value to the free energy of the Grand Canonical Ensemble $\mathcal{F}$, since $\mathcal{I}_E T\equiv \mathcal{F} = \mathcal{M} - T \mathcal{S} - \mathcal{W} q$, 
where $\mathcal{M}$ and $\mathcal{S}$ represent the conserved mass and entropy of the black hole respectively. Making use of a Wick rotation, $t\rightarrow i \tau$, we bring the action into the Euclidean form.
To avoid the conical singularity at the event horizon of the black hole, we impose periodicity of the Euclidean time with a period of $\beta_\tau$, which will be related to the temperature $T$ of the black hole via $ \beta_\tau = 1/T$. As is well known, the Euclidean action has the same functional form as the Hamiltonian of the theory. We thus proceed to recognize the terms $(\partial_\tau \chi)^2/N^2$ as part of the corresponding conjugate momentum of the scalar field.
\begin{equation}
    \label{momentum}
    P\equiv\frac{1}{16\pi}\frac{\partial (\sqrt{|g|} \mathcal{L})}{\partial (\partial_0\Phi)}=\frac{\partial_\tau \chi}{2 N X}\left(b S- 4 r X'X  F_4 \right).
\end{equation}
We also note that since $\partial_\tau^2\chi=0$, in order for the staticity of the metric to hold, $\partial_\tau P=0$ by virtue of the above equation since $X$ is solely dependent on the radial coordinate $r$. This constraint has been assumed throughout the analysis. On the other hand, since $\chi=\chi(\tau)$, we can easily verify that $P$ satisfies the following differential equation:
\begin{equation}
    \label{Pdiff}
    \partial_r(\log P)=\partial_r\left[\log \left(\frac{b S- 4 r X'X  F_4}{2 N X}\right)\right]
\end{equation}
which is independent of its fundamental field $\chi$. This implies that $P$ is not a dynamical degree of freedom but rather acts as a primary constraint on the Euclidean action via the geometric condition of staticity. The dynamics of $P$ could have been found only if we considered a generic $\chi (\tau,r)$ from the start. However, this procedure is unnecessary. In particular, the initial constraint of $\partial^2_t\chi=0$ and $\chi=\chi(t)$ fixes the dynamics of $P$, and as such, there is no need to perform the variation with respect to $P$. 
 This aligns with the fact that  the conservation of the corresponding temporal component of the shift-symmetry Noether current is satisfied solely via geometric considerations, i.e. off-shell. 

Therefore, our Euclidean action now assumes the simple Hamiltonian form of
\begin{equation}
    \label{euclidean2}
        I_{E}=i S=\int d\tau \int dr \left\{N \left[\frac{r h'+h-1}{2}-\frac{r^2+\lambda^2}{2\lambda^2}b S+\frac{ X P^2}{b S-4 r X' X F_4}\right]\right\}-\mathcal{B}.
\end{equation}
The degrees of freedom are $h$, $N$, $X$. Indeed, Hamilton's equations for the doublet $\chi,P$ are satisfied independently of the solution via the very definition of $P$. This verifies the notion that the true degree of freedom in the parity and shift symmetric Beyond Horndeski gravity is the kinetic term $X$. We note that $\mathcal{B}$ is a boundary term to ensure a well-defined variational procedure $\delta \mathcal{I}_E=0$ within the class of fields under consideration. Performing the variation of $I_E$ with respect to $h$, $N$, $X$, we obtain,
\begin{align}
\delta I_E=\int d\tau \int dr \left[ E_h \,\delta h +E_N \,\delta N+ E_X\,\delta X  \right] + \left(\beta_\tau \frac{N}{2}r \delta h\right)\Big|^\infty_{r_h}+\left(N \beta_\tau \frac{P^2 X 4 r X F_4 \delta X}{(bS-4 r X'X F_4)^2}\right)\Big|^\infty_{r_h} -\delta \mathcal{B},
\end{align}
where $E_i$ stands for the Euler-Lagrange equation with respect to the degree of freedom $i$. Inserting the functional form of $P$ via \eqref{momentum} into the differential system ${E_i}=0$, our solution is immediately verified while the integral of \eqref{euclidean2} vanishes on-shell. Indeed, the above equations are satisfied for $N=1$, while the solutions for $h$ and $X$ are as in equations (\ref{eq:h-hom-prop-F}) to (\ref{eq:X-hom-prop}). It is therefore clear that when the field equations hold (on-shell), the Euclidean action $\mathcal{I}_E$ is given by the boundary term $\mathcal{B}_E$.

To ensure a well-defined variational procedure, $\delta \mathcal{I}_E=0$, the variation of the boundary term $\delta \mathcal{B}_E$ must be such that it cancels out the boundary terms \cite{Martinez:2004nb}. 
For convenience we will split the variation of the boundary term into two pieces, one at spatial infinity and one at the horizon $\delta \mathcal{B}_E = \delta \mathcal{B}_E(\infty) - \delta \mathcal{B}_E(r_h)$. At the horizon the variation of the metric function is $ \delta h = -h'(r_h) \delta r_h = - 4\pi T\delta r_h$. 
On the other hand, the variation $\delta X$ is proportional to $\delta q$. Therefore, we can easily verify that
\begin{equation}
    \delta \mathcal{B}=- \frac{\delta M}{T}+\frac{\delta A}{4} +\frac{1}{T}\frac{b r_h}{2}\sum_{n=4}^{\infty}c_{n/2}\left(\frac{\lambda}{\sqrt{2}\sqrt{r_h^2+\lambda^2}}\right)^{n-2}\frac{n-1}{n}(\delta q^n)~,
\end{equation}
where $A=4\pi r_h^2$ is the area of the black hole horizon. Given that we are working within the Grand Canonical Ensemble, the corresponding temperature and the scalar chemical potential of the black hole remain fixed. Recognizing that $I_E=-\mathcal{B}$ on-shell, integration of the above equation provides
\begin{equation}
    I_E=\frac{\mathcal{F}}{T}=\frac{M}{T}-\frac{A}{4}-\frac{1}{T}\left(\frac{b r_h}{2}\sum_{n=4}^{\infty}c_{n/2}\left(\frac{\lambda}{\sqrt{2}\sqrt{r_h^2+\lambda^2}}\right)^{n-2}\frac{n-1}{n}q^{n-1}\right) q.
\end{equation}
We can straightforwardly derive the corresponding thermodynamic quantities from $\mathcal{I}_E=\beta_\tau \mathcal{M} - \mathcal{S} -\beta_\tau\mathcal{W} q$. This gives us the conserved mass and the entropy of the black hole as $\mathcal{M} = M$ and
\begin{equation}\mathcal{S}=\frac{A}{4}=\pi r_h^2,\end{equation} 
respectively. Here, it is evident that the mass is solely determined by $M$, while notably, the entropy remains the same as in the General Relativity case. Additionally, the chemical potential for the scalar field is given by:
\begin{equation}
    \mathcal{W}=\frac{b r_h}{2}\sum_{n=4}^{\infty}c_{n/2}\left(\frac{\lambda}{\sqrt{2}\sqrt{r_h^2+\lambda^2}}\right)^{n-2}\frac{n-1}{n}\left( q^{n-1}\right).
\end{equation} 
The temperature of the black hole can be easily calculated from the periodicity of the Euclidean time,
\begin{equation}
    T=\frac{1}{\beta_\tau}=\frac{1}{4\pi r_h}\left[1+\sum_{n=4}^{\infty}c_{\frac{n}{2}}(n-1)2b\frac{r_h^2}{\lambda^2}\left(\frac{q^2}{2}\frac{1}{1+\frac{r_h^2}{\lambda^2}}\right)^{n/2}\right], \label{tempgeneral}
\end{equation}
where we note that it can vanish at a finite $r_h$ due to the contribution of the primary charge and the fact that $b\sum_{n=4}^{\infty}c_{n/2}<0$. It is worth mentioning that for $q=0$, the temperature will be the Schwarzschild black hole temperature $T=1/(4\pi r_h)$, as expected. Consequently, the first law of thermodynamics in our case reads accordingly
\begin{equation}
    \delta \mathcal{M} = T\delta \mathcal{S} + \mathcal{W} \delta q~, \label{firstlaw}
\end{equation}
which resembles the corresponding first law form of electromagnetically charged black holes and holds \textit{by construction}. Note that since the regular black hole configurations are sourced by a secondary scalar charge, it is clear that $q=q(M)$ and thus the first thermodynamic law will be modified. 

\section{A Special Case} 

In the previous section, we derived the thermodynamics of our solutions within our general framework. However, since our solutions are given in a non-closed form, it will be helpful to consider exact solutions in a closed form to study their properties. The simplest solution occurs when $n=5$. In this section, we will discuss the behavior of this particular solution. For simplicity, we discuss separately the primary and regular classes for $n=5$. It is worth noting that although we examine this simple solution, the above results can be used to study all the possible configurations of interest in a self-consistent manner. 

\subsection{Primary hair solution}
For $n=5$ we can obtain the simple metric function
\begin{equation}
    h(r) = 1-\frac{2 M}{r} +  \frac{\sqrt{2} b \lambda  q^5}{3 r} \left(\frac{r^3}{\left(\lambda ^2+r^2\right)^{3/2}}-1\right)~, \label{chatzi}
\end{equation}
where for simplicity we absorb the constant $c_{5/2}$ into the coupling constant $b$. Therefore, the dimensions of the parameters are $[q]=L^{-1}$, $b=L^{-5}$ and $[\lambda]=L$. 
In this case, the temperature becomes
\begin{equation}
  T = \frac{b \lambda ^3 q^5 r_h}{2 \sqrt{2} \pi  \left(r_h^2+\lambda
 ^2\right){}^{5/2}}+\frac{1}{4 \pi  r_h}.
\end{equation}
%
%
\begin{figure}[t]
    \centering   \includegraphics[width=0.5\textwidth]{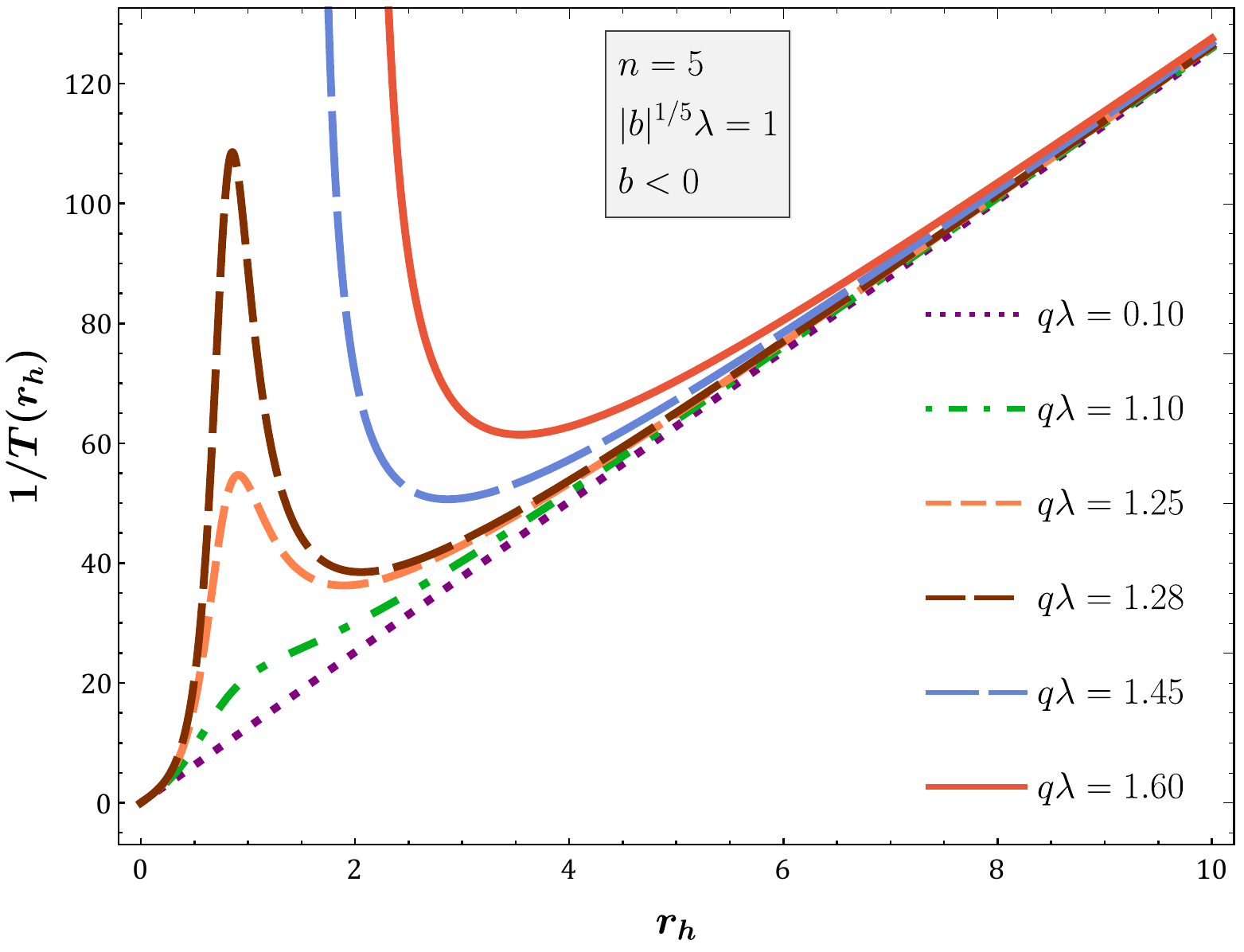}
    \caption{The inverse temperature for a family of primary-hair black hole solutions.}
    \label{fig:tempfree}
\end{figure}
%
%
As can be verified from Fig. {\ref{fig:tempfree}}, the primary charged black holes have an indistinguishable thermal behaviour to Schwarzschild for small primary charge contributions. On the other hand, for larger values of the primary charge, a phase transition occurs to a thermal branch characterized by a positive heat capacity. However, since the temperature does not vanish in this branch, evaporation continues and the black hole becomes thermally unstable again. This implies the existence of a \textit{pseudo-stable} intermediate thermal branch in the black hole configuration space, which is a novel characteristic of these solutions. To stop the runaway evaporation of the hairy black hole, the primary charge needs to be large enough for an interior horizon to form, which is indicated by the cases of $q\lambda=1.45$ and $q\lambda=1.6$. Indeed, in these cases, the temperature reaches a vanishing point which signals the existence of an extremal black hole.  It is also interesting to note that, following the analysis of \cite{Wei:2022dzw,Chatzifotis:2023ioc}, the black hole transitions into a different topological sector of the thermal configuration space with a discrete change in its global topological charge from $-1$ to $0$.

\subsection{Regular Solution}
It is clear that setting $b q^5 = -3 \sqrt{2} M/\lambda $ the black hole becomes regular and the metric function assumes the form
\begin{equation}\label{regm}
    h(r) = 1-\frac{2 M r^2}{\left(\lambda ^2+r^2\right)^{3/2}}~,
\end{equation}
which coincides with the magnetically charged regular ``Bardeen" black hole \cite{Ayon-Beato:2000mjt} of non-linear electrodynamics. 
The electromagnetic Lagrangian supporting the Bardeen black hole assumes a highly non-intuitive form. In our case, however, such a spacetime can be obtained by demanding the regularity of the primary-hair black hole.  Additionally, we observe that the electromagnetic Lagrangian of the Bardeen black hole includes the conserved black hole parameters: the magnetic charge and the mass-to-magnetic-charge ratio. 
As a result, it remains unclear whether these parameters are allowed to vary. Despite this problematic behavior, the first law of thermodynamics for the Bardeen black hole has been investigated in \cite{Ma:2014qma,Lan:2023cvz}. In order for it to hold, either the entropy or the mass of the black hole has to be modified. In our scenario, the first law of thermodynamics is modified to $(1-W\dot{q})\delta \mathcal{M} = T\delta \mathcal{S}$, consistent with the treatment of the first law in \cite{Ma:2014qma} \footnote{Here the dot denotes derivation with respect to the argument.}.
The temperature of the regular black hole and the coefficient of $\delta\mathcal{M}$ are
\begin{equation}
    T(r_h) = \frac{r_h^2-2 \lambda ^2}{4 \pi  \lambda ^2 r_h+4 \pi  r_h^3}~, \qquad \text{and,} \qquad (1-W\dot{q})=1+\frac{3 \lambda ^2 r_h}{25 \left(r_h^2+\lambda ^2\right){}^{3/2}}.
\end{equation}

The coefficient $(1-W\dot{q})$ is positive, and therefore the first thermodynamics law is well-defined.  The temperature reaches a maximum at the radius, $r_* = \lambda\left(\frac{1}{2} \left(\sqrt{57}+7\right)\right)^{1/2} $ resulting in a diverging point in the heat capacity. As the black hole shrinks in size, the temperature increases and reaches its maximum value at $r_*$, causing the heat capacity to be negative, indicating thermodynamical instability. At  $r_h=r_*$ the heat capacity diverges and for $r_h<r_*$ the black holes become colder as they shrink in size, leading to a positive heat capacity until $r_h=\sqrt{2}\lambda$ where the temperature and the heat capacity become null, and evaporation stops.

\section{Conclusions}

In this work, we examine the thermodynamics of primary hair black holes within the framework of shift and parity symmetric Beyond Horndeski gravity. Employing the Euclidean approach, 
we have extracted the first thermodynamic law of the family of solutions under consideration and provided a specific example for elucidation. Notably, the thermal behavior of our black hole presents multiple phase transitions during the process of evaporation, contingent upon the primary charge's value. Specifically, even in scenarios where the black hole lacks an interior horizon to impede evaporation, a pseudo-stable intermediate phase exists wherein the heat capacity becomes positive within a finite range of black-hole sizes. However, since the temperature does not vanish in this regime, the evaporation does not stop and the heat capacity becomes once again negative in the final thermal stages. On the other hand, we verified that when the primary charge induces an interior horizon within the geometry, the evaporation ceases, thereby indicating the existence of eternal primary-charged black holes. Another significant finding is our confirmation of the non-trivial nature of the first thermodynamic law for the regular secondary charged black hole, aligning with the results of \cite{Ma:2014qma}.

To our knowledge, this marks the inaugural exploration into the thermal characteristics of black holes endowed with primary scalar charge, representing a notable milestone in the investigation of these intriguing configurations. It is imperative to underscore that the scrutiny of the associated Euclidean action posed a considerable challenge, requiring rigorous attention. Using our method, one may compute the thermodynamics of lower dimensional black hole solutions that feature a linearly time-dependent scalar field \cite{Bravo-Gaete:2014haa, Ayon-Beato:2004nzi}. Moreover, since a cosmological constant term naturally emerges in our solution, one may study possible Van der Waals analogies by treating it as a thermodynamic quantity, as done in the electrically charged case  \cite{Kubiznak:2012wp}.  


\section{Acknowledgements}
\noindent The research project was supported by the Hellenic Foundation for Research and Innovation (H.F.R.I.) under the “3rd Call for H.F.R.I. Research Projects to support Post-Doctoral Researchers” (Project Number: 7212). A.B. also acknowledges participation in the COST Association Action CA21136 “Addressing observational tensions in cosmology with systematics and fundamental physics (CosmoVerse)”. We are very happy to thank Christos Charmousis, Nick E. Mavromatos and Theodoros Nakas  for useful discussions.

\bibliography{Refs}{}
\bibliographystyle{utphys}

\end{document}